\documentclass[spanish,runningheads]{llncs}

\usepackage[utf8]{inputenc}
\usepackage{babel}
\usepackage{fancyhdr} 
\usepackage{graphicx}
\usepackage{url}

\begin{document}
\title{Hacia una Plataforma de Servicios para Apps Inteligentes en Ciudades Intermedias\thanks{Financiado por CONICET-Argentina mediante proyecto PU-E 2017 ''Plataforma de Servicios para el Desarrollo de Software de Ciudades Inteligentes'', y en colaboración con la Municipalidad de Tandil (Buenos Aires, Argentina)}}
%
%
\author{J. Andres Diaz-Pace\inst{1} \and
Luis Berdun\inst{1} \and
Alejandro Zunino\inst{1} \and
Silvia Schiaffino\inst{1}}
\authorrunning{Diaz-Pace et al.}

\titlerunning{Hacia una Plataforma de Servicios en Ciudades Intermedias}

%
\institute{ISISTAN Research Institute, CONICET and UNICEN University, Tandil Buenos Aires 7000, Argentina 
\url{http://www.isistan.unicen.edu.ar} 
\email{\{andres.diazpace, luis.berdun, alejandro.zunino, silvia.schiaffino\}@isistan.unicen.edu.ar}}
\maketitle              

\thispagestyle{fancy} 

\begin{abstract}
Las ciudades inteligentes (o \textit{smart cities}) constituyen una tendencia en alza en muchas ciudades de Argentina. En particular, las denominadas ciudades intermedias presentan un contexto y requerimientos diferentes a los de las grandes ciudades respecto a ciudades inteligentes. Un aspecto de relevancia es el fomentar el desarrollo de aplicaciones (generalmente para dispositivos móviles) que posibiliten a los ciudadanos aprovechar datos y servicios asociadas normalmente a la ciudad, por ejemplo, en el dominio de movilidad urbana. En este trabajo, se propone una plataforma para ciudades intermedias que provea servicios de ''alto nivel'' y que permita la construcción de aplicaciones de software que consuman dichos servicios. La estrategia centrada en la plataforma apunta a integrar sistemas y fuentes de datos heterogéneos, y proveer servicios ''inteligentes'' a distintas aplicaciones. Ejemplos de estos servicios incluyen: construcción de perfiles de usuario, recomendación de eventos locales, y sensado colaborativo en base a técnicas de data mining, entre otros. En este trabajo, se describe el diseño de esta plataforma (actualmente en progreso), y se discuten experiencias de aplicaciones para movilidad urbana, que están siendo migradas bajo la forma de servicios reusables provistos por la plataforma. 

\keywords{Ciudades intermedias  \and Middleware \and Servicios inteligentes \and Aplicaciones de movilidad urbana.}
\end{abstract}
\section{Introducción}\vspace{-0.2cm}
Una ciudad puede verse como un ecosistema complejo de personas y organizaciones que conviven y trabajan juntos para alcanzar sus objetivos. En los \'ultimos a\~nos, los centros urbanos han tenido un gran impacto en el desarrollo económico y social de los pa\'ises, y este fenómeno ha involucrado no s\'olo a las grandes ciudades sino también a las denominadas \textit{ciudades intermedias} (de aproximadamente 500 mil habitantes) \cite{Bolay:2014}. En el caso de Argentina, esta situación ha sido analizada en varios reportes \cite{Capellan:2016}, con una discusión de temáticas como: movilidad urbana, eficiencia energética, tratamiento de residuos, salud, medioambiente, y comunicación entre gobierno y ciudadanos, que se han trasladado naturalmente al \'ambito de ciudades inteligentes. 

Las ciudades inteligentes (o \textit{smart cities}) constituyen una tendencia en alza en Argentina, que pretende integrar soluciones de las TIC con el fin de mejorar la calidad de vida de sus ciudadanos y su interacción con los responsables del gobierno. El proceso de transformación de una ciudad en una \textit{smart city} normalmente se asocia a grandes centros urbanos, que suelen ser considerados como m\'as proclives para la innovación. Sin embargo, también las ciudades intermedias han cobrado una importancia creciente en lo referido a innovación. Las ciudades intermedias poseen capacidades de generar y consolidar redes que posibiliten la interacción y el flujo de conocimiento entre los diversos actores, tales como las universidades y centros de investigación, las empresas y los gobiernos locales y el sector asociativo \cite{Capellan:2016}. Un ejemplo de dichas ciudades es el caso de Tandil. Un rasgo distintivo de las ciudades intermedias en relación al desarrollo de \textit{smart cities} es su contexto, que normalmente difiere del de una gran ciudad (por ej., Buenos Aires o Córdoba, en Argentina) \cite{Manzano:2015}. Una ciudad intermedia es un centro m\'as f\'acilmente gobernable, y que permite en principio mayor participación ciudadana en el gobierno y la gestión de la ciudad. 

En el caso de ciudades inteligentes, la capacidad de trabajar con datos heterogéneos y de múltiples fuentes es clave \cite{Corbellini:2018}. A esto debe sumarse la capacidad de aplicar t\'ecnicas de minería de datos y de predicción sobre estas fuentes de datos. Desde un punto de vista ingenieril, si bien se han propuesto soluciones específicas para distintos sistemas (por ej., transporte, gestión de residuos, energía o salud), éstas suelen ser ad-hoc y/o focalizadas en conjuntos de datos específicos, con pocas facilidades de reutilización o de interoperabilidad.  La integración de sistemas en una \textit{smart city} es un requerimiento normal en la evolución de dicha ciudad, y requiere contar con una infraestructura (o \textit{middleware}) subyacente que provea una serie de servicios básicos (por ej., soporte al desarrollo de aplicaciones, despliegue, gestión y mantenimiento de aplicaciones). Una estrategia de Ingeniería de Software para esta problemática es la de contar con una arquitectura de referencia \cite{Guessi:2014,Bass:2012} para \textit{smart cities}, que luego pueda derivar en la construcción de una plataforma para desarrollo de aplicaciones.  

Los desafíos antes mencionados llevan a la necesidad de investigar las plataformas de software de servicios para una ciudad intermedia inteligente como un estrato diferenciado, que no responde a las din\'amicas de las grandes ciudades ni a las soluciones de software tradicionales (o comerciales) para ellas. Si bien es factible planificar un ambiente de IoT para una ciudad intermedia, generalmente la infraestructura para IoT requiere de un esfuerzo económico importante a fin poder aprovechar sus beneficios, y no siempre es viable en ciudades del interior de Argentina. Una alternativa a explorar son los dispositivos m\'oviles que, por su ubicuidad en la población y su capacidad para conectarse a Internet, hacen que las apps móviles constituyan un enfoque interesante para configurar una ciudad inteligente. En el marco de una ciudad intermedia es importante lograr una integración de distintas fuentes de información de entrada, y sacar provecho de dicha información para ofrecer servicios útiles a los ciudadanos. En este sentido, se habla de servicios inteligentes cuando se emplean técnicas de minería de datos, procesamiento de lenguaje natural, y an\'alisis de redes sociales, entre otras t\'ecnicas, con el fin de producir alertas, notificaciones, respuestas a preguntas, y recomendaciones, entre otras salidas, que consideren el contexto de un usuario (o grupo de usuarios) o de una aplicación particular. 

En este trabajo, se propone el desarrollo de una plataforma de servicios inteligentes para la ciudad de Tandil, considerada como una ciudad intermedia t\'ipica de Argentina. La plataforma apunta a proveer, progresivamente, servicios  de alto nivel  que permitan (y faciliten) el desarrollo de aplicaciones de software (para la ciudad) en base a dichos servicios. Ejemplos de estos servicios incluyen: construcción de perfiles de usuario, recomendación de eventos locales, y sensado colaborativo, en base a t\'ecnicas de data mining y procesamiento paralelo distribuido. Se argumenta que esta estrategia centrada en plataforma promover\'a un proceso paulatino de transformación de Tandil en una ciudad m\'as conectada e inteligente. Como dominio inicial para la plataforma, se está trabajando sobre aplicaciones vinculadas a transporte y movilidad urbana. 

El resto del artículo se encuentra estructurado en 5 partes. En la Sección 2 se provee un marco conceptual sobre \textit{smart cities}, con un \'enfasis en la categoría de ciudades intermedias. En la Sección 3 se discute la propuesta de plataforma y su arquitectura. En la Sección 4 se presentan algunas experiencias con aplicaciones de transporte que, mediante un proceso de refactorización, están actualmente siendo incorporadas como servicios a la plataforma. La Sección 5 está abocada a trabajos relacionados. Finalmente, la Sección 6 presenta las conclusiones y plantea l\'ineas de trabajo futuro.  

\section{Marco Conceptual}\vspace{-0.2cm}
Se han propuesto varias definiciones de \textit{smart city} en la literatura \cite{Chourabi:2012,Neirotti:2015,Anthopoulos:2015,Belissent:2013}, algunas m\'as orientadas hacia los servicios que se proveen a los ciudadanos, otras hacia la gestión eficiente de recursos urbanos, y otras concentradas en sociedad y economía.
En el presente trabajo, una ciudad inteligente puede definirse como un ambiente urbano caracterizado por el uso de tecnología para: i) facilitar la coordinación o integración de sub-sistemas urbanos; y ii) mejorar la experiencia y calidad de vida de las personas dentro del ambiente urbano. El primer aspecto conlleva un monitoreo continuo de (sub-)sistemas por parte del ente de gobierno de la ciudad, y la disponibilización de los datos recolectados hacia los ciudadanos. El segundo aspecto implica un aprovechamiento ''inteligente'' de la información y la provisión de servicios con valor agregado, que permitan una interacción fluida entre el gobierno y los ciudadanos. La característica de inteligencia no es menor en este marco, ya que el mero hecho de poner a disposición conjuntos de datos en forma electrónica no implica necesariamente un uso efectivo de dichos datos por los ciudadanos. Una característica relacionada es la de usabilidad de los servicios e información provista, por ejemplo, mediante tecnologías Web y de dispositivos móviles.

Las tecnologías más comúnmente empleadas en \textit{smart cities} \cite{Santana:2017} incluyen: Internet of Things (IoT), Big Data, \textit{cloud computing}, y los denominados sistemas ciber-f\'isicos, sumado a la computación ubicua y orientación a servicios \cite{Khatoun:2016}. En particular, en el caso de las ciudades inteligentes, la capacidad de trabajar con datos heterogéneos y provenientes de m\'ultiples fuentes es clave. Uno de los dominios donde es posible aprovechar distintas tecnologías y fuentes de información es la movilidad y el transporte urbano.\vspace{-0.4cm}
\subsection{Ciudades intermedias}\vspace{-0.2cm}

Las denominadas ''ciudades intermedias'' constituyen un fenómeno creciente en Latinoamérica, y en particular, en Argentina \cite{Manzano:2015}. Por su propia naturaleza y din\'amica, diferente a la de las grandes urbes, las ciudades intermedias tienen la posibilidad de llevar a cabo proyectos de desarrollo territorial sostenibles y de mejorar su función de servicios e infraestructura. La dinámica de una ciudad intermedia suele implicar el establecimiento de relaciones con otros núcleos urbanos y con el campo. Adicionalmente, una ciudad intermedia, por su propia escala, tiene la posibilidad de llevar a cabo proyectos de desarrollo territorial sostenibles y de fomentar emprendimientos (por ej., turismo). Existen varias ciudades intermedias en Argentina con agendas relacionadas con \textit{smart cities}. Una de estas ciudades es Tandil, y se toma como referencia para el presente trabajo. Los entornos urbanos de las ciudades intermedias plantean, para los responsables de la gestión municipal, demandas de: eficiencia, desarrollo sostenible, calidad de vida y gestión de los recursos. En este escenario, la aplicación de las TIC aparece como una respuesta concreta a la urbanización no planificada y a la necesidad de orientar esta expansión a mejorar la calidad de vida de las personas~\cite{Finquelievich:2016}.

Tandil \footnote{\url{http://www.tandil.gov.ar}} es una ciudad intermedia de la provincia de Buenos Aires, que ha crecido de forma continua y heterogénea durante los últimos años. La población actual ronda los 150.000 habitantes, y la ciudad se extiende de forma irregular en un valle con una superficie aproximada de 50 km$^2$. El crecimiento urbano de la ultima década ha sido disperso y espontáneo, por distintas razones, y no siempre ha respondido a un plan u organización que haya considerado la disposición de las infraestructuras de transporte y servicios. La principal actividad productiva de Tandil es la agricultura intensiva (trigo, soja, maíz, girasol) y la ganadería, aunque también se desarrollan otras actividades como el turismo y la actividad metal\'urgica. La presencia de la Universidad Nacional del Centro (UNICEN) y de diversos centros de investigación ha permitido más recientemente el desarrollo de actividades científico-tecnológicas, y fomentado la radicación de varias empresas dedicadas a las TIC. Esto ha impulsado con fuerza el sector de software y servicios informáticos que hoy emplea alrededor de 1200 personas \cite{Finquelievich:2016}.

El gobierno municipal de Tandil ha incorporado la utilización de TIC en diversas ramas, desde nuevos servicios a los ciudadanos (por ej., acceso a trámites vía Internet, mayor conectividad, Internet gratuito en espacios públicos) a mecanismos de acceso a la información pública. Desde 2015, se desarrolla el denominado Sistema Único de Movilidad Ordenada (SUMO) que utiliza una tarjeta única para el estacionamiento medido, el boleto electrónico para buses, y próximamente un sistema de bicicletas públicas. También se ha implementado el uso de las redes sociales como medio de comunicación con el ciudadano. 

No obstante, el ecosistema de Tandil y la introducción de las TIC plantea ciertos desafíos para la ciudad. Un primer desafío es lograr un modelo participativo y colaborativo de los ciudadanos en relación a la gestión municipal, por ejemplo, a través de canales de redes sociales o de otros mecanismos de participación ciudadana. El ciudadano ya no es un ente pasivo que internaliza servicios en su vida, sino un consumidor activo e informado, que se integra a un proceso de interacción bi-direccional de la información entre el cliente y el proveedor de los servicios. Un segundo desafío es el de fomentar nuevas oportunidades de negocio para las empresas de la región, y particularmente para las empresas basadas en conocimiento como son las empresas de software. Un tercer desafío tiene que ver con lograr una mejor articulación del territorio, ya que la presencia territorial del Estado no es todavía homogénea (por ej., en ciertos barrios). 

Por otro lado, se debe considerar que si bien la inversión en tecnología en la ciudad ha sido importante en los últimos años (por ej., en lo referido a semaforización y cámaras de vigilancia), no se prevé una incorporación masiva de sensores (IoT) en la ciudad, en parte por el costo económico asociado y porque los ciudadanos no perciben (todavía) beneficios en el corto plazo de dichos sensores para su quehacer diario. En este contexto, es m\'as factible aprovechar los dispositivos móviles de los ciudadanos para una estrategia de ciudad inteligente. 
\vspace{-0.2cm}
\subsection{Movilidad urbana}\vspace{-0.2cm}
La movilidad urbana se refiere a la gestión los medios de transporte dentro de la ciudad y a los costos, en tiempo y dinero, que implican  para los ciudadanos trasladarse de un lugar a otro para realizar sus actividades diarias. Una \textit{smart city} tiene que estar comprometida con un uso eficiente y multi-modal del transporte, fomentando el transporte p\'ublico y aquellas opciones con menor impacto sobre la polución del ambiente. En las ciudades intermedias, el transporte tanto público (por ej., buses) como privado (por ej., automóviles) es utilizado por gran parte de los ciudadanos, incluso muchos de ellos dependen de su correcto funcionamiento. Este tipo de cuestiones, casi rutinarias, de las personas que se ven afectadas por la experiencia que perciben de los servicios de transporte son factores determinantes a la hora de plantear alternativas para mejorarlo. Dentro de las características para mejorar los servicios se distingue, por ejemplo, el pago electrónico mediante el uso de tarjetas, lo que permite contar con información específica de los viajes de cada usuario. 

Movilidad urbana es una de las aplicaciones típicas de una \textit{smart city}, por ejemplo, mediante apps para transporte público o provisión de rutinas personalizadas para un usuario. Para el diseño de estas aplicaciones (generalmente con soporte de dispositivos m\'oviles) se debe proveer de información de valor a los usuarios a fin de lograr elecciones inteligentes para el viaje~\cite{Cairns:2005}. Al mismo tiempo, las empresas de transporte deben comprometerse a mejorar la calidad de los servicios prestados, la cuál también es influenciada por la cantidad y calidad de información que es suministrada por parte de los usuarios.

\begin{figure}
\centering \includegraphics[width=0.62\columnwidth]{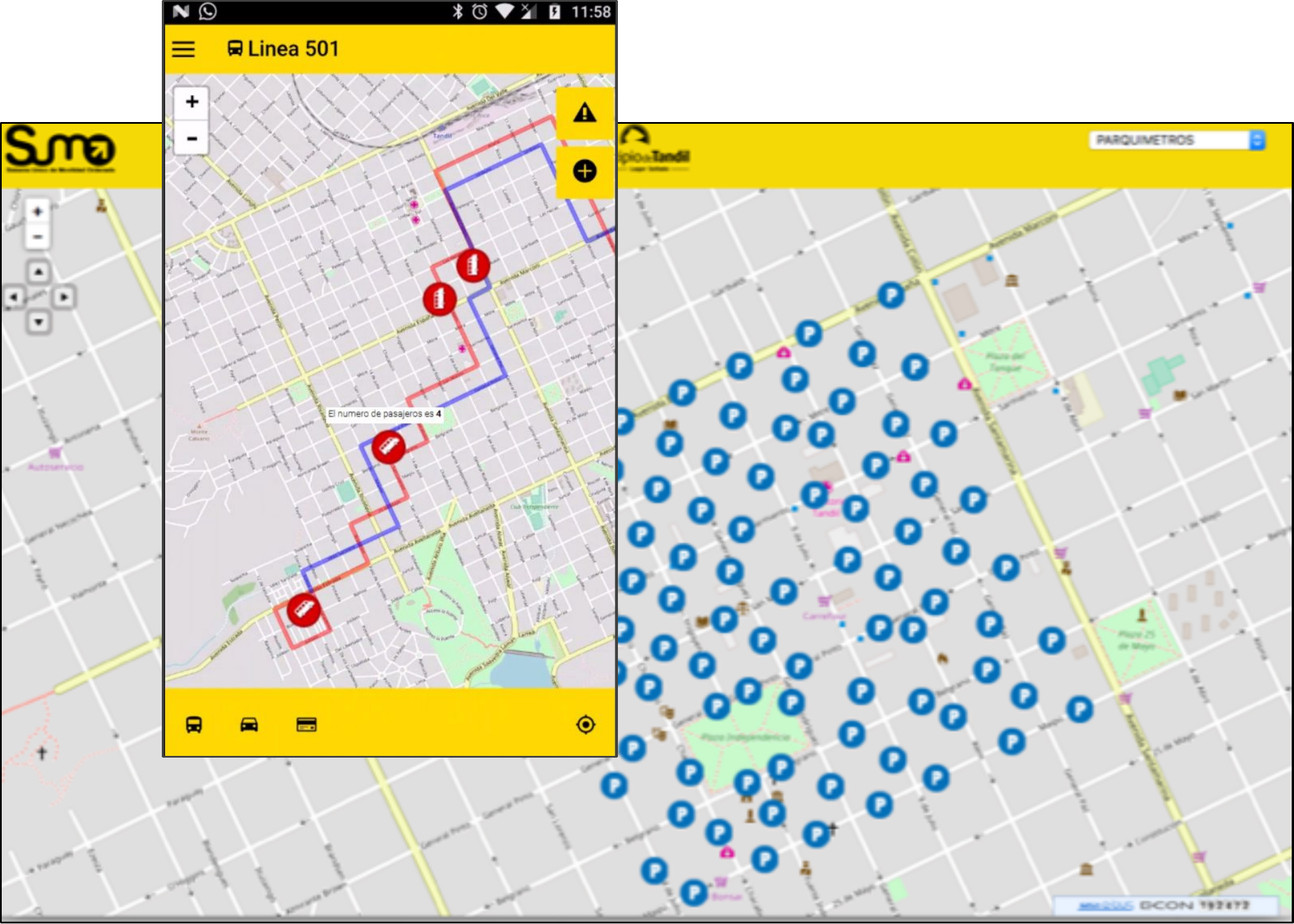}
\vspace{-0.3cm}
\caption{Ejemplo de sistema SUMO para movilidad urbana en Tandil.} \label{fig-ejemplo-sumo}\vspace{-0.5cm}
\end{figure}

En la actualidad, Tandil cuenta con el sistema SUMO para algunos aspectos de movilidad urbana. Este sistema es operado mediante tarjetas, y es accesible a través de la Web y de dispositivos  m\'oviles, como se muestra en la Figura~\ref{fig-ejemplo-sumo}. Este tipo de sistemas presenta ventajas como: i) la no utilización de dinero en efectivo al subir a un bus, ii) la consulta de saldo por parte de los usuarios, y iii) la visualización de los recorridos de las líneas de buses de la ciudad (en tiempo real), entre otros. Esto brinda un mejor servicio al ciudadano en el sentido de que puede manejar mejor sus tiempos a la hora de tomar un bus. El sistema SUMO también permite gestionar, mediante las mismas tarjetas y unidades de registro (parquímetros) por cuadra, el estacionamiento de vehículos en un sector predeterminado (centro) de la ciudad. El software del sistema SUMO es provisto por una empresa privada. Sin embargo, estas aplicaciones y fuentes de datos no está actualmente diseñadas para integrarse con otras fuentes de información u otros sistemas. Este fue uno de los disparadores para nuestro trabajo sobre una plataforma integradora.
\vspace{-0.5cm}

Toda ciudad que se considere inteligente debería promover (dentro de sus necesidades y capacidades específicas) un transporte p\'ublico eficiente, aprovechando la información que los mismos ciudadanos puedan brindar (ya sea, voluntariamente, o a través de sensores tecnológicos) para asegurar la velocidad y disponibilidad (con tiempos coherentes) de los medios de transporte. A su vez, el ciudadano debe poder estar informado para decidir mejor sobre sus opciones de movilidad, y también poder informar sobre problemas en el servicio.

\section{Enfoque centrado en Plataforma}\vspace{-0.2cm}
Dadas las características particulares de las ciudades intermedias con planes de volverse \textit{smart cities} en Argentina, en este trabajo se propone un enfoque de automatización centrado en tres bloques básicos: i) la construcción de una plataforma de servicios a nivel municipal, ii) la provisión de un conjunto de servicios montados sobre dicha plataforma, y iii) el desarrollo de un conjunto de apps que consuman los servicios de la plataforma. En este enfoque, el desarrollo de la plataforma toma preponderancia (por sobre las apps), ya que constituye el corazón del sistema en lo referido a integración de datos de fuentes heterogéneas, y provisión de capacidades inteligentes para disponibilizar datos hacia las apps. 
 
Además de los ciudadanos (o usuarios finales), la plataforma posee dos destinatarios principales: los organismos  públicos, y los desarrolladores de aplicaciones. Por un lado, los organismos públicos (por ej., la misma municipalidad de Tandil) buscan mejorar la calidad de vida y servicios públicos a ofrecer a los ciudadanos mediante soluciones de \textit{smart cities}, y para ello se enfrentan a cuestiones de: costos, tiempo, dependencia de soluciones propietarias, y aspectos de privacidad, entre otros. Por otro lado, los desarrolladores de aplicaciones para \textit{smart cities} se enfrentan a problemas de: falta de estándares, difícil acceso a datos públicos, y largos procesos de integración de sistemas. En este sentido, la plataforma busca facilitar y potenciar la construcción de aplicaciones, con foco en dispositivos móviles, para desarrolladores y empresas de la región.

La provisión de servicios inteligentes es un diferenciador del enfoque, ya que permite valerse de técnicas de Inteligencia Artificial y Machine Learning para proveer información con valor agregado, en función de la combinación de distintas fuentes de datos. Por ejemplo, ante problemáticas diarias de una ciudad como: los conductores que desean saber donde hay un lugar de estacionamiento disponible, o los ciudadanos que desean saber cuando llegará el siguiente bus, pueden emplearse algoritmos de predicción o de recomendación. Esto implica no considerar cada aplicación o proveedor de información como un ''silo''. Por ejemplo, el sistema SUMO actual funciona correctamente pero se centra solamente sobre el estacionamiento medido y la gestión del transporte p\'ublico de buses. Un desarrollador de una app móvil podría tomar esta información, y en vez de simplemente mostrársela al usuario, podría combinarla con información municipal para determinar lugares donde está prohibido estacionar, y reusar un servicio de ruteo y geolocalización, e intentar inferir lugares libres para una ventana de tiempo dada. Con estos datos, la predicción podría incluir un ranking (o recomendación) de los lugares con mayor probabilidad de estar desocupados.

\begin{figure}
\centering \includegraphics[width=0.81\columnwidth]{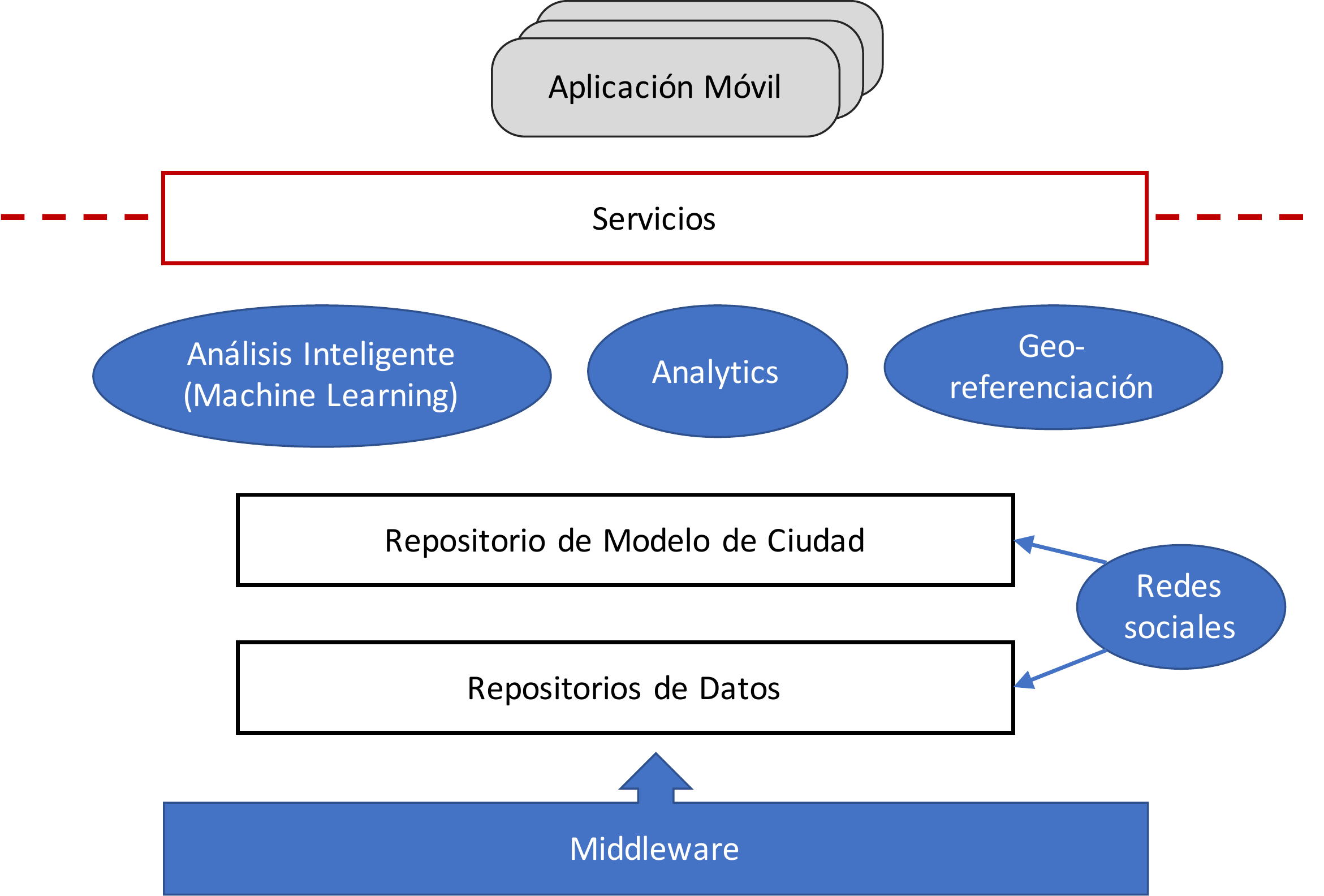}
\vspace{-0.3cm}
\caption{Esquema de arquitectura de referencia para ciudad intermedia \textit{smart-city}.} \label{fig-plataforma}
\end{figure}

En lo que respecta a la construcción de la plataforma y la progresiva evolución de sus capacidades (en el tiempo), se considera beneficiosa la definición de una \emph{arquitectura de referencia}~\cite{Bass:2012}, que partiendo de modelos existentes en la literatura, provea las capas principales de abstracción y grupos funcionales para una \textit{smart city}. La Figura \ref{fig-plataforma} presenta una organización inicial de la arquitectura de referencia para la plataforma, derivada de \cite{Santana:2017}, adaptada al contexto de una ciudad intermedia (como Tandil), y a las características de procesamiento y análisis inteligente de datos buscadas por nuestro enfoque. Una arquitectura de referencia (de software) define los principales bloques funcionales, la asignación de dichas funcionalidades a componentes, y sus patrones de interacción, para un determinado dominio o conjunto de sistemas. Cabe destacar que cuando se construye (o materializa) un sistema determinado a partir de una arquitectura de referencia, no necesariamente deben utilizarse todos los bloques sino que puede ensamblarse el sistema a partir de un sub-conjunto de ellos. 

\vspace{-0.5cm}

\vspace{-0.5cm}

Los principales elementos de la plataforma propuesta son:
\begin{itemize}
  \item \textbf{Middleware.} Representa el nivel base de recepción y envío de mensajes y eventos de la plataforma, ya sea para entidades externas a la plataforma como para entidades internas. Se asume un mecanismo de comunicación desacoplado, bajo un esquema publicador/subscriptor, que brinde flexibilidad a la hora de integrar nuevos sistemas y también portabilidad. 
  \item \textbf{Repositorios de Datos.} Este nivel captura las diferentes fuentes de información que provee la plataforma, que pueden ser: archivos, bases de datos relacionales, bases de datos NoSQL, y streams de datos, entre otras.
  \item \textbf{Repositorios con Modelo de Ciudad.} Constituye un modelo de información opcional, que facilita la manipulación, integración, y entendimiento de los datos de la plataforma (nivel anterior) por parte de los otros elementos.
  \item \textbf{Análisis Inteligente. }Se refiere a la utilizaci\'on de t\'ecnicas de Inteligencia Artificial para procesar información (generalmente, en formatos no estructurados), con el prop\'osito de identificar patrones y realizar predicciones. Estas capacidades de an\'alisis est\'an orientadas a aplicaciones de \textit{smart cities} m\'as que a aplicaciones generales de Inteligencia Artificial.
  \item \textbf{Analytics.} Provee capacidades para análisis descriptivos (por ej., visualizaciones),  y predictivos de la información de los repositorios. Generalmente, funciona en tándem con el análisis inteligente, pero puede ser opcional.
  \item \textbf{Geo-referenciación. } Normalmente, las aplicaciones requieren un soporte de mapas y geo-localización, que permitan situar al usuario o a eventos, o bien integrar información disponible en cierta región de la ciudad.
  \item \textbf{Redes Sociales.} Se consideran una fuente particular de información bi-direccional (entre los ciudadanos y la plataforma, y vice-versa), que a menudo puede alimentar las funcionalidades de análisis inteligente y de analytics.
  \item \textbf{Servicios.} Los elementos previos son internos de la plataforma. La interacción con aplicaciones y el aprovechamiento de la funcionalidad anterior se realiza mediante una capa de servicios. Estos servicios generalmente están expuestos como servicios Web (por ej., con tecnología REST).
\end{itemize}

Para la materialización de esta arquitectura de referencia en una plataforma de software se plantea una estrategia ''conducida por ejemplos'', a partir de la definición de una arquitectura base y de una serie de aplicaciones existentes. Inicialmente, no es necesario que dichas aplicaciones estén construidas sobre la plataforma, ya que la idea es re-factorizarlas en función de servicios que podría proveer la plataforma, y en ese ejercicio progresivamente construir y mejorar los servicios de la plataforma. En otras palabras, en una primera fase, la estrategia de desarrollo consiste en que la infraestructura de servicios emerja a partir de las necesidades concretas de aplicaciones existentes. En una segunda fase, se espera que las aplicaciones se construyan directamente sobre la plataforma, aprovechando servicios existentes en ella.

\section{Experiencias: Asistencia para Movilidad Urbana}\vspace{-0.2cm}

En esta sección, se presentan una serie de aplicaciones para movilidad urbana en Tandil que, a partir de una aplicación tradicional cliente-servidor, han ido evolucionado y mejorando sus servicios, y a la vez contribuyendo a un primer prototipo de la plataforma proyectada.

\vspace{-0.3cm}

\subsubsection*{AsisTAn.} 
\begin{figure}\vspace{-0.5cm}
\centering \includegraphics[width=0.91\columnwidth]{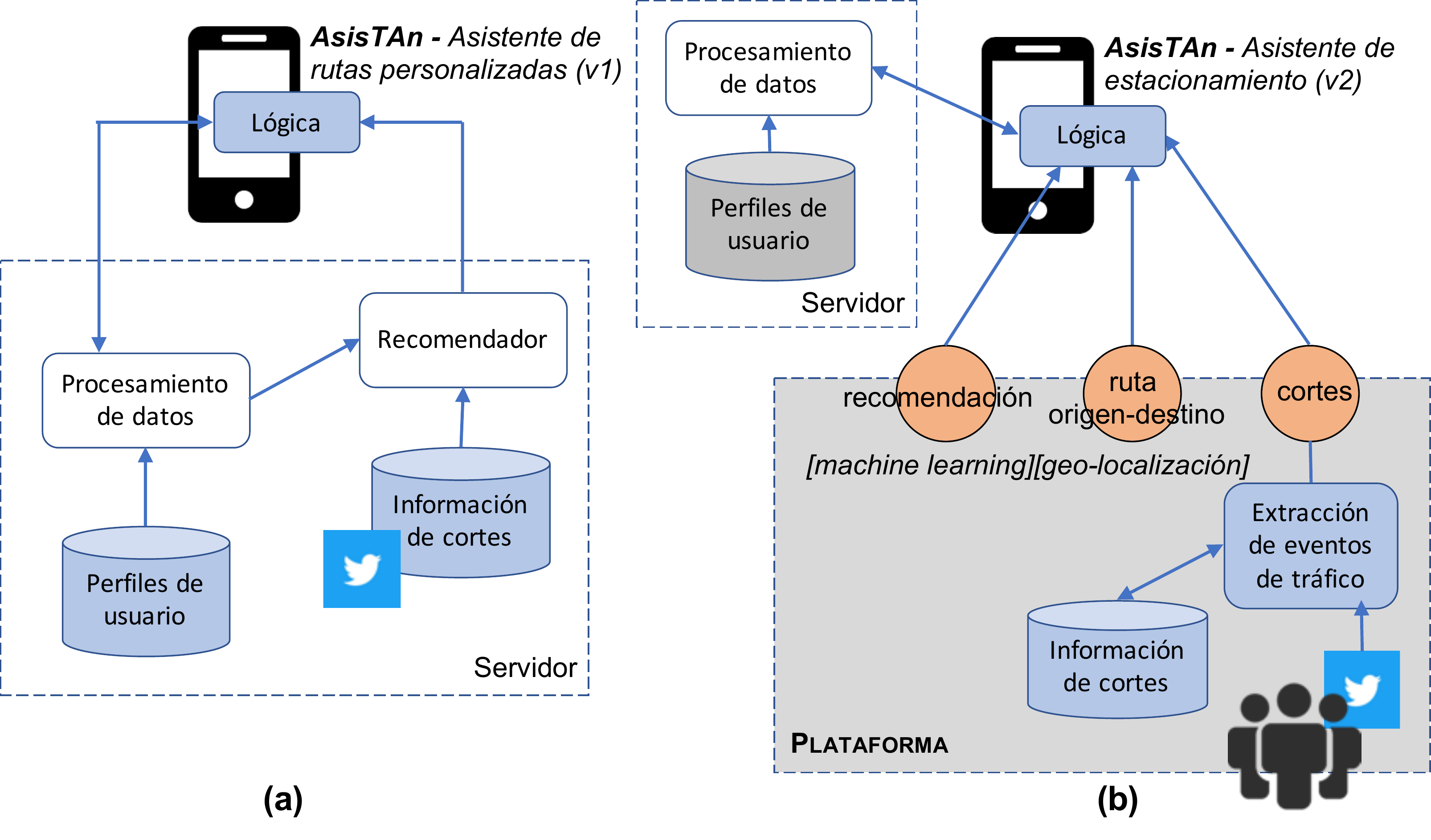}
\vspace{-0.3cm}
\caption{Transformación de una aplicación ad-hoc (o standalone) (a) en una aplicación que provea los servicios de la plataforma} \label{fig-transformacion-asistan}
\vspace{-0.6cm}
\end{figure}

Esta aplicación móvil  fue inicialmente construida para brindar asistencia personalizada a un conductor en sus rutinas cotidianas \cite{Caimmi:2016,DCristofaro:2016}. Dicha asistencia se focaliza en eventos de tránsito que pueden afectar dichas rutinas. La aplicación aprende las rutinas del conductor, y sumándole información de eventos de tránsito, le provee rutas alternativas para llegar a su destino sin inconvenientes, en caso de que su rutina se vea afectada por dichos eventos. El dispositivo m\'ovil permite capturar la información del usuario que es utilizada para poder aprender, de forma no intrusiva, su rutina. 
El asistente provee una visualización de las rutinas aprendidas y los recorridos de las mismas en un mapa. 
Los experimentos y simulaciones realizadas sobre la herramienta generaron resultados prometedores, en términos de precisión en el reconocimiento de actividades rutinarias, proveyendo recomendaciones satisfactorias para diferentes perfiles de usuario. La Figura \ref{fig-transformacion-asistan} ilustra cómo el diseño inicial (a) que fue desarrollado en forma ad-hoc, es decir, sin utilizar servicios de plataforma, puede evolucionarse hacia un nuevo diseño (b) que aprovecha los servicios de la plataforma. 

La nueva versión de \textit{AsisTAn} sigue utilizando un servidor donde se encuentran almacenados los perfiles de usuario, pero re-implementa su funcionalidad de recomendación de rutas en base a tres servicios de la plataforma (marcados por círculos en la figura). Estos servicios (refactorizados) tienen que ver con: la obtención de caminos (no personalizados) entre un origen y un destino dados, la extracción de información publicada por ciudadanos en Twitter sobre problemas de tránsito, y el algoritmo base de recomendación. Estos servicios se corresponden con las capacidades de: geo-referenciación, repositorios de datos y redes sociales, y análisis inteligente, respectivamente. Se decidió no migrar los perfiles de usuario a la plataforma por cuestiones de confidencialidad. 

La Figura \ref{fig-plataforma-evolucion} muestra el estado actual del desarrollo de la plataforma, donde se están integrando dos nuevas aplicaciones: \textit{Manwe} y \textit{EsTAcionAR}, para reporte de eventos de tráfico y de estacionamiento de veh\'iculos, respectivamente.

\vspace{-0.25cm}\subsubsection*{Manwe.} Esta aplicación es una mejora a la funcionalidad de extracción de información de tránsito a partir de Twitter que existía en \textit{AsisTAn}. \textit{Manwe} \cite{Caimmi:2016} encapsula a nivel plataforma un servicio de clasificación, análisis y geolocalización de tweets.  Se combinan técnicas de Machine Learning y de Procesamiento de Lenguaje Natural para procesar los tweets y detectar incidentes de tránsito. \textit{Manwe} se encarga de registrar en una base de datos los incidentes reportados en Twitter. Los incidentes detectados son geolocalizados en un mapa de la ciudad. A cada incidente se le asigna una duración, para indicarle al usuario su vigencia. Notar que la información de cortes también se puede alimentar de información de los sitios institucionales de la municipalidad de Tandil.

\vspace{-0.35cm}

\subsubsection*{EsTAcioNAR.} 
\begin{figure}
\vspace{-0.5cm}
\centering \includegraphics[width=0.91\columnwidth]{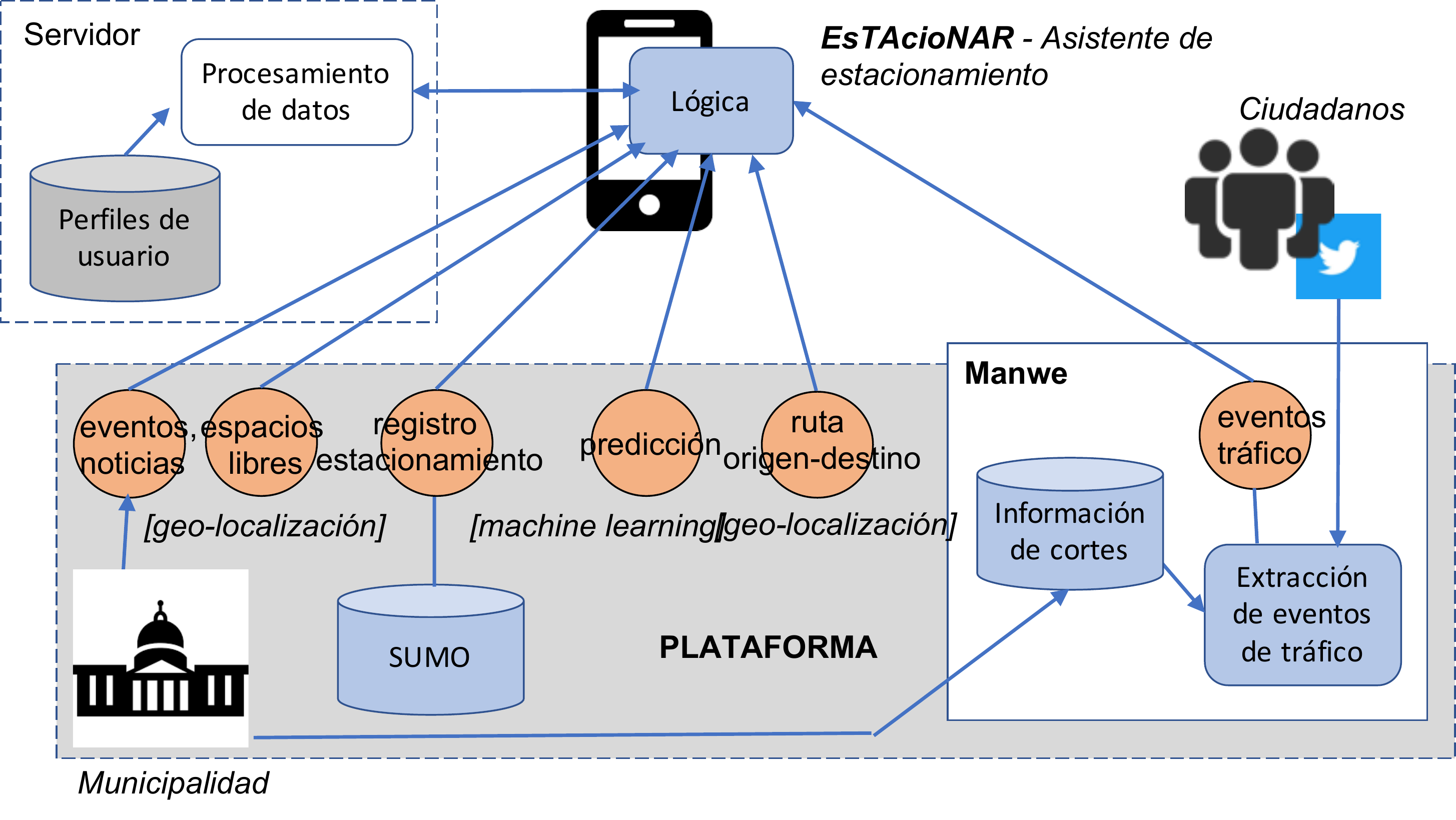}
\vspace{-0.4cm}
\caption{Evolución de la plataforma, incorporando nuevas aplicaciones y servicios} \label{fig-plataforma-evolucion}
\vspace{-0.5cm}
\end{figure}

Esta aplicación móvil busca asistir a los conductores a encontrar un espacio libre de estacionamiento dentro del perímetro de estacionamiento medido (de SUMO). Si bien el sistema actual de estacionamiento permite obtener información básica sobre el número total de automóviles estacionados por cuadra, esta información no está accesible al ciudadano mediante una app. Tampoco se brinda asistencia para indicar al conductor si puede encontrar un espacio libre en una determinada cuadra. La información hoy capturada por SUMO es parcial en este sentido, ya que no considera factores como: las cocheras por cuadra, las zonas prohibidas para estacionamiento vehicular, ni el vehículo particular que desea estacionar y su ruta hasta el centro de la ciudad. La estrategia adoptada para construir la aplicación consiste en aprovechar fuentes heterogéneas información para proveer una predicción razonable de un lugar libre. Para ello, se combinan varios servicios. En primer lugar, se parte de la información sobre registro de estacionamientos provista por SUMO (parquímetros). A esto se le suman servicios para identificar avisos del municipio de Tandil, y para determinar los posibles lugares libres por cuadra (en base a un análisis geográfico de cada cuadra en particular). El algoritmo de predicción está basado en series de tiempo y Machine Learning \cite{Zheng:2015}. Los servicios de la Figura  \ref{fig-plataforma-evolucion} se corresponden a las capacidades de: geo-referenciación, repositorios de datos, análisis inteligente, y redes sociales, de la plataforma.

Un punto relevante de \textit{EsTAcioNAR} es que reutiliza los desarrollos previos de \textit{Manwe} y de otros servicios ya disponibles en la plataforma. Adicionalmente, se desea notar que el ''valor agregado'' de la aplicación respecto a sus capacidades de predicción implica una combinación de servicios m\'as que un desarrollo ad-hoc para el problema de estacionamiento. En consecuencia, el esfuerzo de desarrollo de la app es relativamente bajo, en comparación con un enfoque de tipo ''silo''.

Se desea notar que varios de los datos urbanos a almacenar en los repositorios de la plataforma están generalmente administrados por las autoridades municipales. En el caso de Tandil, estos datos están regidos por la implementación de una ordenanza municipal orientada a datos abiertos. En este sentido, se ha establecido un proyecto de cooperación y trabajo conjunto, a fin de poder incorporar dichos datos en la plataforma.  

\vspace{-0.2cm}
\section{Trabajos Relacionados}\vspace{-0.2cm}
Un relevamiento de tecnologías, plataformas, y algunas aplicaciones para \textit{smart cities} se presenta en \cite{dasilva:2013,Santana:2017}, y en base a este an\'alisis, los autores proponen una arquitectura inicial de referencia que sirve de base para el presente trabajo. A partir del an\'alisis de las plataformas existentes, se observa que las principales actividades de una plataforma están orientadas a controlar el ciclo de vida de los datos de la ciudad en lo que se refiere a: i)~recolectar datos con una red de sensores wifi, ii)~administrar los datos en la plataforma, iii)~procesar los datos utilizando un modelo de ciudad, y iv)~compartir los datos procesados permitiendo el acceso externo. Por otro lado, se destacan aspectos no funcionales de la plataforma tales como: reutilización de componentes ya desarrollados, flexibilidad para evolucionar en el tiempo, escalabilidad (ante incrementos en la demanda), performance en tiempo real (en caso de ser necesario), y usabilidad por parte de usuarios y aplicaciones, entre otros aspectos de calidad.
 
Una instancia típica de plataforma de \textit{smart city} es SmartSantander~\cite{sanchez:2014}, una infraestructura experimental centrada en la ciudad de Santander (España) y con instalaciones en otras ciudades de Europa. Santander puede considerarse como una ciudad intermedia en el contexto europeo. SmartSantander procesa una gran variedad de información, condiciones del tr\'afico, temperatura, emisiones de CO2, humedad y luminosidad, mediante un conglomerado de sensores. El diseño de SmartSantander puede no ser fácilmente trasladable a otras ciudades, debido a su modelo de ciudad y la estrecha relación con la plataforma de IoT. Una plataforma que trabaja de forma similar a SmartSantander es PadovaSmartCity~\cite{zanella:2014}, que utiliza IoT para crear una red de sensores en la ciudad de Padua (Italia). En relación a esto último, la plataforma europea para ciudades inteligentes (EPIC)~\cite{ballon:2011} propone un middleware para el uso y administración de la red de sensores wireless (WSN), contemplando cuestiones de heterogeneidad, interoperabilidad, escalabilidad, extensibilidad y configurabilidad en WSN. Por otro lado, Civitas~\cite{Villanueva:2013} es un middleware orientado a objetos para manejar distintos dispositivos y sensores, y que adicionalmente incorpora una capa para procesar la información recolectada mediante técnicas de Big Data. SmartSantander, PadovaSmartCity y Civitas pueden considerarse como desarrollos basados en la tecnología de IoT, con un énfasis fuerte en la incorporación de hardware embebido y sensores en el espacio urbano. Esta característica, si bien novedosa, las diferencia del contexto y expectativas de las ciudades intermedias en Argentina. Por otro lado, las plataformas no incluyen aspectos de procesamiento inteligente de la información y personalización, las cuáles son responsabilidad de las aplicaciones que pueden desarrollarse sobre estas plataformas.

Recientemente, en \cite{delesposte:2017} se discute la provisión de plataformas gen\'ericas para \textit{smart cities}, en contraposición a soluciones ad-hoc para ciudades específicas. Los autores discuten tres factores que contribuyen a desarrollar soluciones particulares. Un primer factor tiene que ver con la necesidad de cada ciudad de satisfacer requerimientos no funcionales diferentes (por ej., seguridad, escalabilidad, disponibilidad), y la evolución de las aplicaciones en el tiempo, en lo que refiere a necesidades de integración con distintas fuentes de datos o cumplimiento de normativas. El segundo factor se refiere a la falta de herramientas para evaluar el impacto de las tecnologías de una \textit{smart city} en los ciudadanos, o para comparar distintas aplicaciones o plataformas. El tercer factor se vincula al bajo desarrollo de las soluciones \textit{open-source} respecto a soluciones comerciales y a problemas de interoperabilidad entre proveedores de soluciones. En este contexto, los autores hacen una propuesta \textit{open-source} de plataforma basada en micro-servicios llamada IntersCity, en base a la arquitectura de referencia de~\cite{dasilva:2013}. IntersCity, en principio, puede ser adaptada a diferentes dominios, tales como transporte p\'ublico, seguridad y ambiente, y puede ser consumida por aplicaciones Web y móviles. Otra plataforma genérica \textit{open-source} que ha logrado algunos casos de éxito es Sentilo\footnote{\url{http://www.sentilo.io}}. De forma similar a otras soluciones, el objetivo está en proveer mecanismos para almacenamiento de datos y su posterior análisis, por ejemplo, mediante sistemas de reglas para generación de alarmas. Este tipo de plataformas comparten varias de las ideas desarrolladas en nuestro enfoque, aunque se centran primordialmente en ambientes IoT, y solo abordan m\'inimamente el procesamiento inteligente de los datos. 
 
Existen en el mercado algunas soluciones comerciales con plataformas para \textit{smart cities}, como las provistas por IBM\footnote{\url{https://www.ibm.com/smarterplanet/us/en/smarter_cities/overview}} o Living PlanIT\footnote{\url{http://www.living-planit.com}}, que están orientadas a grandes urbes más que a ciudades intermedias. Si bien estas soluciones pueden funcionar en ciudades de distinto tamaño, sus componentes están abocados al control de dispositivos, y no tanto al procesamiento inteligente de datos.

Respecto a movilidad urbana, algunas empresas están comenzando a ofrecer soluciones para diferentes problemas, generalmente de ciudades grandes. Por ejemplo, Smart Parking\footnote{\url{https://www.smartparking.com}} diseña, desarrolla y administra tecnologías de estacionamiento inteligente.  En particular, posee servicios para encontrar lugares libres, ya sea dentro de la ciudad como fuera de ella. Los servicios se basan en el uso de sensores, cámaras y análisis inteligente de datos.  Por otro lado, aplicaciones como GoogleMaps utilizan datos de geolocalización provistos por dispositivos m\'oviles para analizar la velocidad del tráfico en cualquier instante de tiempo. Tras adquirir Waze en 2013, GoogleMaps puede además incorporar incidentes de tráfico reportados por los usuarios, tales como: accidentes, zonas en construcción o reparación.  Utilizando esta información GoogleMaps puede sugerir, por ejemplo, la ruta más rápida hacia o desde el trabajo.

\vspace{-0.3cm}
\section{Conclusiones y Perspectiva}\vspace{-0.1cm}

Las \textit{smart cities} han dado origen a avances en soluciones para obtener datos mediante sensado de diferentes aspectos de las ciudades. Si bien existen soluciones ad-hoc y comerciales en este sentido, existen necesidades particulares de las ciudades intermedias que motivan nuevas formas de explotar la información disponible, en base a un aprovechamiento, por ejemplo, de capacidades de sensado participativo en dispositivos móviles de los ciudadanos. En este trabajo se presentaron las principales consideraciones de diseño de una plataforma para \textit{smart cities}, acotada a necesidades de ciudades intermedias, y particularmente al caso de Tandil. Además, se describió una experiencia de aplicaciones de movilidad urbana que involucra tanto la plataforma como algunos algoritmos de análisis de datos, y los desafíos que este escenario plantea.

Existen trabajos que sugieren que los usuarios de transporte poseen necesidades de información con ''valor agregado'', tales como funciones personalizadas de avisos tempranos e información multi-modal sobre el viaje \cite{wang:2014}. Actualmente, si bien existen ciertas soluciones comerciales y \textit{open-source}, dichas soluciones no consideran los hábitos de cada usuario, sino que actúan de la misma forma para todos los usuarios. Por otro lado, estas soluciones no están diseñadas necesariamente para ciudades intermedias, donde el foco est\'a puesto en un desarrollo y gestión urbana eficiente.
En este sentido el trabajo propuesto permite proveer a los desarrolladores de apps de servicios reusables con fines de personalización y asistencia en la movilidad urbana. Poder conocer el estado actual de un parquímetro, o inferir la ocupación de un bus en pleno recorrido, permiten agregar una capacidad innovadora al desarrollo de una app para movilidad urbana. Actualmente, se está trabajando en el desarrollo de servicios para la plataforma que brinden información de congestionamiento en las principales calles a partir del análisis de vídeo de las cámaras de seguridad instaladas en la ciudad. 

Respecto a la implementación  de los módulos de servicios en la plataforma, que se espera poder ofrecer a los desarrolladores de apps, se ha observado que la separación entre dichos servicios y las fuentes de datos sobre las cuáles debieran operar no siempre es clara. En algunas ocasiones, resulta más práctico que la plataforma ofrezca los datos directamente a los desarrolladores de apps (consumidores), y que ellos construyan sus propios servicios. Eventualmente, algunos de estos servicios podrían luego volcarse a la plataforma. En consecuencia, se está evaluando contemplar un modelo colaborativo de desarrollo.

Como próximos pasos, se cuenta con un relevamiento inicial y orientativo sobre objetivos del gobierno municipal de Tandil en lo referido a una \textit{smart city}, y también sobre preocupaciones de los ciudadanos. Adicionalmente a la movilidad urbana, estos aspectos tienen que ver con: estado de las calles, animales callejeros (denuncia, control, búsqueda de hogares), control de espacios verdes (baldíos, si hay basura, corte de pastizales), acceso a espacios recreativos (mantenimiento, limpieza, objetos rotos: farolas, bancos, juegos), recolección de residuos (no pasan los recolectores, basura arrojada en espacios públicos), y turismo (circuitos para visitantes, ubicación en la ciudad, opciones de servicios al turista). Para la plataforma en si, se planea brindar un mejor soporte para aplicaciones móviles, y a corto plazo seleccionar un \textit{middleware open-source} y basado en micro-servicios.


\vspace{-0.35cm}
%
%
%
%
\bibliographystyle{splncs04}
\bibliography{bib_sample}

\end{document}